\documentclass{ws-procs975x65}

\newcommand{\p}{\partial}
\newcommand{\Lagr}{\mathcal{L}}
\newcommand{\I}{\mathcal{I}}
\newcommand{\J}{\mathcal{J}}

\begin{document}



\title{GRAVITATION AS A VACUUM NONLINEAR ELECTRODYNAMICS EFFECT}

\author{ALEXANDER A. CHERNITSKII
}

\address{A. Friedmann Laboratory for Theoretical Physics, 
St.-Petersburg, Russia\\
and\\
State University of Engineering and Economics, 
Marata str. 27, St.-Petersburg, Russia, 191002\\
\email{AAChernitskii@eltech.ru}}


\begin{abstract}
Effective Riemann space effect of vacuum nonlinear electrodynamics is considered in the context of theory for unified gravitation and electromagnetism. The electromagnetic four-vector potential in the scope of Born-Infeld nonlinear electrodynamics model is considered as the unified field. The electromagnetic interaction appears naturally in the first perturbation order by the small field of distant material objects. The gravitational interaction appears naturally in the appropriate second order.
In this case the effective metric components contain the corresponding energy-momentum tensor components for quick-oscillating electromagnetic field of the distant objects.
\end{abstract}

\bodymatter

\section{Introduction}

The problem of unification for all interactions of material particles remains one of the
most important problems in modern theoretical physics. Particularly, unification for
the two known long-range interactions, viz., electromagnetism and gravitation, must be considered as
a priority problem in theoretical investigation of nature.

The approach to this problem considered here is connected with a consistent application of the idea of nonlinear
local unified field theory.

Spatially localized solutions in this theory correspond to solitary material particles.
These solutions can be designated particle solutions.
The many-particle world configuration corresponds to a complicated many-particle world solution.
Any many-particle solution contains the appropriate particle solutions in the following sense.

Each particle solution has at least ten free parameters for space-time rotation and shift.
Because of the nonlinearity, a sum of particle solutions is not a solution for the model.
But we can consider the free parameters of particle solutions to be weakly time-dependent.
This method is well known in the theory of nonlinear dynamics.
A sum of particle solutions with time-dependent free parameters can be considered
as an initial approximation to a many-particle solution. The time dependence of the free parameters
of the particle solutions corresponds to the interaction between the particles.

This method, applied to a suitable model, must give electromagnetic and gravitational interactions for
the case when the interacting particles are sufficiently distant from each other.

For the first time, this approach to the problem of unification of the gravitational and electromagnetic interactions
appeared in the context of some nonlinear electrodynamics model \cite{Chernitskii1992}.
Later on, the approach was developed for another nonlinear electrodynamic model (Born-Infeld) \cite{Chernitskii1999}.

A distinguishing characteristic of this approach is that the gravitational interaction must appear through
an effective Riemannian space for propagation of particle under consideration.
This effective Riemannian space is induced by the electromagnetic field of distant particles.

\section{Model field equations}

A variational principle of the model considered here is similar to the one proposed by A.S. Eddington  in the context of
general relativity ideas \cite{Eddington1924} and investigated by A.~Einstein in the context of his unified field
theory \cite{Einstein1923dE}.
Afterwards, M.~Born and L.~Infeld used it in the context of nonlinear electrodynamics \cite{BornInfeld1934a}.

The model set of equations is \cite{Chernitskii1998a}
\begin{equation}
\label{35357121}
\frac{\p}{\p x^\mu}\,\sqrt{|g|}\; f^{\mu\nu} {}={} 0
\;\;,\quad\quad
%
f^{\mu\nu} {}\equiv {}
\frac{1}{\Lagr}\,\left(F^{\mu\nu}  {}-{}
\frac{\chi^2}{2}\,\J\,\varepsilon^{\mu\nu\sigma\rho}\,F_{\sigma\rho}\right)
,\;
\end{equation}
where
$\Lagr {}\equiv{}   \sqrt{|\,1 {}-{}  \chi^2\,\I  {}-{}  \chi^4\,\J^2\,|}$,
\mbox{$\I {}\equiv{}  F_{\mu\nu}\,F^{\nu\mu}/2$},
\mbox{$\J {}\equiv{}  \varepsilon_{\mu\nu\sigma\rho}\, F^{\mu\nu} F^{\sigma\rho}/8$},
$\varepsilon_{0123} \equiv \sqrt{|g|}$,
$\varepsilon^{0123}  = - 1/\sqrt{|g|}$, $F_{\mu\nu}\equiv\p_\mu A_\nu - \p_\nu A_\mu$, 
$A_\mu$ is the electromagnetic potential.

The model energy-momentum tensor is
\begin{equation}
\label{71416389}
T^\mu_{.\nu} \equiv  \left[f^{\mu\rho}\,F_{\nu\rho}  {}-{}
\chi^{-2}\,\left(\Lagr {}-{} 1\right)\,\delta^\mu_\nu\right]/4\pi
\;\;.
\end{equation}

With the definitions
$E_i \equiv F_{i0}$, $B^i \equiv -\varepsilon^{0ijk}\, F_{jk}/2$,
$F_{ij} = \varepsilon_{0ijk}\, B^k$,
$D^i \equiv  f^{0i}$, $H_i \equiv \varepsilon_{0ijk}\, f^{jk}/2$,
$f^{ij} = -\varepsilon^{0ijk}\, H_k$ (Latin indexes take the values $1,2,3$),
 the model equations can be written as
nonlinear equations for the electromagnetic field (see also \cite{Chernitskii2004a}).
This system has the characteristic equation \cite{Chernitskii1998b}
\begin{equation}
\label{CharEq}
\tilde{g}^{\mu\nu}\,\frac{\p \Phi}{\p x^\mu}\,\frac{\p \Phi}{\p x^\nu}=0
\;,
\end{equation}
where $\Phi (x^\mu)=0$ is the equation of a characteristic surface,
and the induced metric $\tilde{g}^{\mu\nu}$ has the following very notable form,
which is specific for the model under consideration:
\begin{equation}
\label{73144072}
\tilde{g}^{\mu\nu} \equiv g^{\mu\nu} - 4\pi\,\chi^2\,T^{\mu\nu}
\;\;.
\end{equation}
Here $T^{\mu\nu}$ is the energy-momentum tensor defined in (\ref{71416389}).

\section{Electromagnetic and gravitational interactions}

The model to be considered is an unusual electrodynamics not only because the appropriate equations are
nonlinear. This model does not contain the postulated trajectory equation for charged particle in external
electromagnetic field, i.e., the postulated electromagnetic interaction. This interaction appears naturally as a
manifestation of the nonlinearity of the model.

The electromagnetic interaction appears in this model as an electromagnetic force acting on a massive charged particle \cite{Chernitskii1999}
and a moment of force acting on a particle with an electric or magnetic dipole moment and spin \cite{Chernitskii2006a}.
The appropriate dynamical equations follow from integral conservation laws for the field energy-momentum and
angular momentum, respectively (for details see \cite{Chernitskii1999,Chernitskii2006a}).
These obtained equations, characterizing the electromagnetic interaction, have the corresponding classical form.

The force and the moment of force contain the electromagnetic field of distant particles in the first power.
Thus we can say that the electromagnetic interaction appear in the first order in the small field of
distant particles.

An explanation of the gravitational interaction in the scope of this model based on the effective Riemannian space
with the metric $\tilde{g}^{\mu\nu}$ (\ref{73144072}) induced by the electromagnetic field.
According to the general method stated in the Introduction, distant particles modify the propagation conditions
for particle under consideration by means of this effective Riemannian space induced
by the field of distant particles. The effective metric includes the electromagnetic field components in
even powers. Thus we can say that the gravitational interaction appear in the second order in the small field of
the distant particles.

The cause of the gravitational interaction in this approach is the energy density of the distant particle
field. But to have the real behaviour of the gravitational potential, i.e. $1/r$, we must take into account
a quick-oscillating part of the distant particle field with an electromagnetic wave background.
In this case, an averaging can give the necessary behaviour of the energy density, $1/r$
\cite{Chernitskii2002b,Chernitskii2006b}.

For additional details see also \cite{Chernitskii2006c}.


\section{Conclusion}

Thus the present approach, based on a consistent application of the idea
of a nonlinear local unified four-vector field, can really unify electromagnetism and gravitation.


\begin{thebibliography}{10}

\bibitem{Chernitskii1992}
A.~A. Chernitskii, {\em Theoret. and Math. Phys.} {\bf 90}, 260 (1992).

\bibitem{Chernitskii1999}
A.~A. Chernitskii, {\em J. High Energy Phys.} {\bf 12}, Paper 10 (1999),
  hep-th/9911093.

\bibitem{Eddington1924}
A.~Eddington, {\em The Mathematical Theory of Relativity} (Cambridge, 1924).

\bibitem{Einstein1923dE}
A.~Einstein, Zur allgemeinen \uppercase{R}elativit\"atstheorie, in {\em
  Sitzungsber. preuss. \uppercase{A}kad. \uppercase{W}iss., phys.-math.\/}
  (Kluwer Academic Publishers, 1923).

\bibitem{BornInfeld1934a}
M.~Born and L.~Infeld, {\em Proc. Roy. Soc. A} {\bf 144}, 425 (1934).

\bibitem{Chernitskii1998a}
A.~A. Chernitskii, {\em Helv. Phys. Acta} {\bf 71}, 274 (1998), hep-th/9705075.

\bibitem{Chernitskii2004a}
A.~A. Chernitskii, Born-\uppercase{I}nfeld equations, in {\em Encyclopedia of
  \uppercase{N}onlinear \uppercase{S}cience\/},  ed. A.~Scott (Routledge, New
  York and London, 2004).
\newblock hep-th/0509087.

\bibitem{Chernitskii1998b}
A.~A. Chernitskii, {\em J. High Energy Phys.} {\bf 11}, Paper 15 (1998),
  hep-th/9809175.

\bibitem{Chernitskii2006a}
A.~A. Chernitskii, Mass, spin, charge, and magnetic moment for electromagnetic
  particle, in {\em X\uppercase{I} \uppercase{A}dvanced \uppercase{R}esearch
  \uppercase{W}orkshop on \uppercase{H}igh \uppercase{E}nergy \uppercase{S}pin
  \uppercase{P}hysics
  (\uppercase{D}\uppercase{U}\uppercase{B}\uppercase{N}\uppercase{A}-\uppercase{S}\uppercase{P}\uppercase{I}\uppercase{N}-05) \uppercase{P}roceedings\/},
  eds. A.~V. Efremov and S.~V. Goloskokov (JINR, Dubna, 2006).
\newblock hep-th/0603040.

\bibitem{Chernitskii2002b}
A.~A. Chernitskii, {\em Gravitation \& Cosmology} {\bf 8}, Supplement, 157
  (2002), gr-qc/0211034.

\bibitem{Chernitskii2006b}
A.~A. Chernitskii, hep-th/0602079.

\bibitem{Chernitskii2006c}
A.~A. Chernitskii, {\em Gravitation \& Cosmology} {\bf 12}, 130 (2006),
  hep-th/0609204.

\end{thebibliography}

\end{document}